# Magnetoresistive effect in graphene nanoribbon due to magnetic field induced band gap modulation

S. Bala Kumar[a)], M. B. A. Jalil
Department of Electrical and Computer Engineering, National University of Singapore, Singapore, 117576,

S. G. Tan
Data Storage Institute, (A*STAR) Agency for Science, Technology and Research, DSI Building, 5 Engineering Drive 1, Singapore 117608

Gengchiau Liang[b)]
Department of Electrical and Computer Engineering, National University of Singapore, Singapore, 117576,

[*]Corresponding author. E-mail: a) brajahari@gmail.com and b) elelg@nus.edu.sg

## Abstract

The electronic properties of armchair graphene nanoribbons (AGNRs) can be significantly modified from semiconducting to metallic states, by applying a uniform perpendicular magnetic field (B-field). Here, we theoretically study the bandgap modulation induced by a perpendicular B-field. The applied B-field causes the lowest conduction subband and the top-most valence subband to move closer to one another to form the n=0 Landau level. We exploit this effect to realize a device relevant MR modulation. Unlike in conventional spin-valves, this intrinsic MR effect is realized without the use of any ferromagnetic leads. The AGNRs with number of dimers, $N_a$=3p+1 [p=1,2,3,…] show the most promising behavior for MR applications, with large conductance modulation and hence, high MR ratio at the optimal source-drain bias. However, the MR is suppressed at higher temperature due to the spread of the Fermi function distribution. We also investigate the importance of the source-drain bias in optimizing the MR. Lastly, we show that edge roughness of AGNRs has the unexpected effect of improving the magnetic sensitivity of the device and thus increasing the MR ratio.

**I. Introduction**

The growth of research activities surrounding graphene-based materials and their applications increases considerably in the last few years due to their unique physical properties for promising device applications, such as the extremely high carrier mobilities,[1,2] fractional quantum Hall effects,[3,4] chiral tunneling (the Klein paradox of relativistic quantum mechanics)[5,6], and bandgap opening of bilayer graphene under the electrostatic field.[7,8] Several promising device applications have thus been demonstrated and proposed.[9-12] Furthermore, another advantage of graphene is the possibility to pattern the graphene nanoribbons (GNRs) to induce a bandgap via quantum confinement, and achieve a semiconductor-like behavior. This versatility has spawned various studies on the transport,[16-18] magnetic,[19-25] and optical properties[24,25] of GNRs.

In addition to the quantum confinement due to the finite size in the transverse direction, an external magnetic field also causes an effective confinement, which further constrains the electron motion in the GNR. At sufficiently large B-field, the cyclotron radius of electron motion becomes smaller than the GNR width, resulting in the formation of Landau levels (LLs). One of the interesting properties of graphene and GNRs is its anomalous zeroth Landau level (n=0 LL), which resides permanently at the Dirac points even as the magnetic field varies.[27-29] This is unlike the case of other materials, e.g. two-dimensional electron gases, where the LLs are created only within the conduction band or the valence band, and these levels shift as the applied magnetic field changes.[30] At smaller B-field, i.e. when the cyclotron radius of electron motion is larger than the GNR width, a different effect is observed. It was found that the bandgap of armchair GNR (AGNR) is reduced with application of the B-field. This is because the lowest conduction subband and the highest valence subband shift closer to one another (they would ultimately meet to form the n=0 LL). Due to this bandgap reduction, the electronic properties of AGNRs can be significantly modified by the application of a B-field.[24,25,31,32]

Recently, magnetoresistive (MR) effects have been explored experimentally[33,34] and theoretically[35-38] in graphene based structures. Using ab-initio calculations, Kim *et al.*[38] predicted a very large MR in spin-valve (SV) devices with zig-zag GNR (ZGNR) channel, owing to the unique spin and orbital symmetry of the ZGNRs.[38] However, using tight-binding (TB) approximation, Bery *et al.*[36] predicted very small MR in a 2D graphene-based SV, due to the weak dependence of the graphene conductivity on the electronic parameters of the ferromagnetic (FM) contacts. Similarly Saffarzadeh *et al.*[37] also showed that the planar FM/GNR/FM junction with zig-zag (armchair) interfaces exhibits a high (low) MR ratio. The above transport calculations have been supplemented by experimental works. For instance, Hill *et al.*[34] have experimentally observed a 10% MR ratio in a GNR based SV device, where a 200nm GNR was connected to NiFe contacts.

In all these previous studies, the MR effect was induced by the change in the relative magnetic orientations of the left and right FM contacts. However, in this work, we will focus on an intrinsic MR property in AGNRs, which arises from the decrease of the bandgap under an applied B-field, as a result of the subband energy shifts in the formation of the n=0 LL. We investigate the MR effect in AGNRs, and calculate its dependence on temperature ($T$) and source-drain bias voltage ($V_{SD}$). We find that utilizing the intrinsic effect, the AGNR can provide a high MR ratio under optimal $V_{SD}$ even without FM contacts. Additionally, we also study the effect of edge roughness (ER) on the electronic transport and magnetic sensitivity of the AGNR. This is in view of the fact that it is still an experimental challenge to fabricate precise edge nanostructure in GNRs and avoid undesired ER. We find that ER can induce an increase in the bandgap of AGNRs. This in turn, results in a rather unexpected improvement of the magnetic sensitivity and the MR of our device. Note that, unlike typical MR effect which is related to the spin asymmetric scattering, the MR effect reported in this paper is solely the result of the B-field induced bandgap modification.

## II. Methodology

Figure 1(a) shows the schematic structure of an AGNR. We employ the π-orbital tight binding model to investigate the electronic structure of an AGNR under a uniform perpendicular B-field. The real space π-orbital tight binding Hamiltonian[31,39] of a GNR is

$$H = \sum_{n} V_n a_n^+ a_n - \sum_{n,m} t_{n,m} a_n^+ a_{m'} \tag{1}$$

where $V_n$=0 is the onsite energy at site $n$, and $t_{n,m}$ (-3 eV) is the hopping energy between two bonded atoms $m$ and $n$. In the presence of a B-field, the Hamiltonian, H(B), is modified based on Peierls phase approximation[24,40]. The B-field of $\vec{B} = (0,0,B_Z)$ induces a vector potential of $\vec{A} = (-B_Z y, 0, 0)$ which satisfies $\nabla \times \vec{A} = \vec{B}$. The hopping energy, $t_{n,m}$ acquires a phase, i.e. $t_{n,m}(B) = t_{n,m}(0) \exp\left( i \frac{q}{h} \int_{l_m}^{l_n} A(B) d\vec{l} \right)$, where $l_{n(m)}$ is the coordinate of atom $n$ ($m$), and $t_{n,m}$ (0) is the hopping energy under zero B-field. The electron transport behaviors of the AGNR are studied using the non-equilibrium Green's function (NEGF) formalism[41]. In the NEGF formalism, the zero temperature conductance, g(E,B) across the GNR is given by:

$$g(E,B) = g_0 Tr[\Gamma_S(E,B)G^r(E,B)\Gamma_D(E,B)G^r(E,B)^+], \tag{2}$$

where $G^r(E,B) = [EI - H(B) - \sum_S(E,B) - \sum_D(E,B)]^{-1}$ is the retarded Green's function of the GNR channel, $\sum_{S(D)}(E,B)$ is the self energy of the source (drain) leads, $\Gamma_{S(D)}(E,B) = i[\sum_{S(D)}(E,B) - \sum^+_{S(D)}(E,B)]$, and $g_0 = q^2/h$. To obtain the current, semi-infinite normal metal leads are assumed at both ends, whose density of state is taken to be constant around the Fermi level region and is set to 0.03/eV/atom/spin. The normalized current density across the structure, J, is computed as follows:

$$J = \frac{1}{L_y}\frac{2\pi}{q}\int_{-\infty}^{+\infty}\Delta f(E)g(E,B)dE \quad , \tag{3}$$

where $\Delta f(E) = f_S(E) - f_D(E)$, $f_{S(D)}(E) = \left[1 + \exp\left(\frac{E - \mu_{S(D)}}{kT}\right)\right]^{-1}$, $\mu_{S(D)} = E_F + (-)V_{SD}/2$,

$E_F$ is the Fermi level to be set at zero (intrinsic level of AGNRs), $L_y$ is the width of AGNR and $V_{SD}$ is the applied bias voltage across the structure. Finally, the MR ratio is determined as follows:

$$MR = \frac{J(B) - J(0)}{J(0)} \times 100\% = [J(B)/J(0) - 1] \times 100\% \quad , \tag{4}$$

where J(B) and J(0) is the current density at finite and zero B-fields, respectively.

## III. Result and discussion

## III.1 Ideal AGNR

Figure 1(b) illustrates the energy dispersion relations and zero temperature conductance of a typical AGNR structure with (sold line) and without (dashed line) B-field, respectively. It can be observed that the subband edges are shifted under an applied B-field. $E_{Y1}$ is defined as the band-edge energy of the first subband in the absence of a B-field. At energy levels below $E_{Y1}$, i.e. $|E| < E_{Y1}$, the electron transmission is zero. When B-field is applied, the lowest subband is shifted to a lower energy level, $E_{Y1}$'. Therefore in the presence of magnetic field, the transmission will be finite for the range of energy in between $E_{Y1}$ and $E_{Y1}$' ($E_{Y1} < E < E_{Y1}$') – see the circled region in Fig. 1(b). This indicates the possibility of modulating the electronic transmission in AGNRs via application of a B-field and by a proper selection of the Fermi level and the operating bias. Due to the electronic sensitivity of AGNR with varying B-field, AGNRs show high (low) conductance at high (low) B-field, resulting in a MR effect (Eq. 4).

Next, we investigate the electronic sensitivity of AGNR width with varying B-field. A previous study has shown that AGNRs can be divided into three different families,[42] namely, AGNR1, AGNR2, and AGNR3, based on the total number of rows in the AGNR (i.e., GNR width), $N_a = (3p+1)/(3p)/(3p-1)$, where p is an integer. As shown in the inset of Fig. 2, AGNR1 always show the largest variation in $E_g$ under a fixed B-field compared to AGNR2/AGNR3 in the same width range (same *p* value). At high B-field the quantum confinement of electrons caused by the finite spatial width becomes less significant than that due to the applied B-field, and thus AGNRs of similar widths will have similar bandgaps. At B=0, AGNR1 has the highest $E_g$[42]. Hence, at higher B-field the $E_g$ of AGNR1 drops more drastically compared to that of AGNR2/AGNR3, implying higher magnetic sensitivity of AGNR1. As shown in Fig.2, amongst AGNRs with widths in both ranges of 7-8 nm (p=20) and 10-11 nm (p=29), AGNR1 in either range shows a steeper drop with increasing B-field. Since the higher sensitivity of AGNR1 to B-field indicates greater suitability for MR applications, we will focus on AGNR1 with $L_y$ of 15.4 nm (p=41) in the following analysis.

Figure 3(a) shows the g(E,B) profile when semi-infinite metallic contacts are connected to the AGNR as the source and drain, and a finite bias of $V_{SD}$=150 mV is applied across the device. The inset of Fig. 3(a) plots the product g(E,B)×Δf(E) as a function of E. The temperature is set at T=100 K, such that Δf(E) ≈ 1 within the transmission window of -e$V_{SD}$/2 < E < e$V_{SD}$/2. It is clear that the area below the g(E,B)×Δf(E) curve, which gives a measure of the current J, is larger when a finite B-field is applied, resulting in a monotonic increase of J with applied B-field [see Fig. 3(b)]. Based on Eq. 3, the increase in J can be attributed to the shift of the $E_{Y1}$ to lower energy, thus causing more g(E,B) peaks to fall within the transmission window where Δf approaches 1. The increase in J(B) with increasing B also translates to an increase in MR (following Eq. 4), as shown in Fig. 3(c). Furthermore, we investigate the temperature dependence on J(B) and MR of AGNRs, and found that as temperature increases from T=100 K to T=300 K, J(B) increases but MR decreases for a fixed B-field. The former trend can be understood by the broadening of Δf(E) due to increasing temperature. As a result, the overlap between Δf(E) and g(E,B) increases and thus, J(B) increases. However, it is interesting to note that the increase of J(0) is more than twice larger than the increase of J(B=15T) as the temperature increases. This is because at larger B-field, g(E,B) is shifted closer to the peak of Δf(E) between E=$\mu_S$ and E=$\mu_D$, i.e. E=0. As T increases, Δf(E) broadens and thus the magnitude of Δf(E) closer to the peak decreases while the tail of Δf(E) increases. This results in a relatively smaller increase of J(B>0) compared to J(0) which is mainly dominated by the tails of Δf(E), and thus causes the MR ratio to degrade with increasing T, as shown in Fig. 3(c).

Next, we study the dependence of $V_{SD}$ on MR effects in AGNR devices. Fig. 4 shows MR variation with increasing $V_{SD}$ at $T=0$ K under $B=5$ T. At zero temperature, the $\Delta f(E)$ is a rectangular function and thus transmission occurs only within the energy range of $-eV_{SD}/2 < E < +eV_{SD}/2$. When $V_{SD} < E_g$, the transmission window falls within the band-gap region, and thus the tunneling current from the source to drain is almost zero for a sufficiently long channel. When $V_{SD} > E_g$, the transmission window becomes larger, and encompasses the lowest subband. This leads to an increase in $J(0)$, and thus a smaller fractional increase in current when a B-field is applied. This in turn translates to a lower MR following Eq. 4. The decrease in MR occurs in an oscillatory manner, which reflects the discrete profile of the conductance peaks of $g(E,B)$, as shown in Fig. 3(a). Interestingly, when $V_{SD}$ is large enough, the MR ratio may even be negative. This is because, at higher $V_{SD}$, part of the contribution to $J(0)$ comes from the electrons occupying the higher subbands (2[nd], 3[rd], and so on). Unlike the lowest subband which shifts towards $E=0$ with an applied B-field, the higher subbands move towards higher E. Hence, at certain B-fields, they will move out of the transmission window, c.f., Fig. 1(b), resulting in decrease of $J(B)$ and negative MR. Referring to the inset of Fig. 4, the oscillations in MR are washed out by the broadening of the Fermi function as the temperature increases. At $T=100$ K, the MR can still be tuned to reach a peak of close to 100% at $V_{SD} \approx 75$ mV. But at higher temperatures (i.e. $T = 200$ K and 300 K), however, the MR monotonically decreases to zero with increasing $V_{SD}$. The results illustrate the dual role of the source-drain bias $V_{SD}$ in our proposed device – i) to maximize the MR ratio by ensuring that $V_{SD}$ coincides with the peak MR value, and ii) to ensure a sufficiently large current, so as to drive the whole circuit for device applications. Therefore, it is important to optimize the $V_{SD}$, such that a high MR can be obtained with sufficient current density.

**III.2 Influence of edge-roughness on the MR of AGNR.**

In a more realistic device, imperfections like edge-roughness are unavoidable. Thus it is important to study the performance of the device in the presence of atomic disorders at the edges. The schematic structure of a smooth AGNR, and an AGNR with edge-roughness are shown in the inset of Fig. 5(c). The ER is modeled by the following principle. The carbon dimers along the top and bottom edges were randomly removed according to the percentage of ER assumed. A similar number of dimers will be added randomly above and below the top and bottom row of the GNRs. The ER is quantified as the percentage of carbon atoms which are dislocated at the edges. Based on the atomic geometry with the required level of ER, the new Hamiltonian H(B) is generated. In this work, for each ER level, five samples are generated randomly. Then, the computed J and MR are averaged over the five samples to obtain the overall behavior.

Fig. 5(a) and (b) respectively show the current density (J) and MR of the individual ER geometric configurations (plot symbol '+') and the average value (denoted by the solid line). In general, under a magnetic field of B = 5T, the average J decreases with increasing ER while the average MR increases with increasing ER. To explain these trends, we plot g(E,B) for devices with smooth (ER=0%) and rough (ER=25%) edges. When B=0T, we found that the conductance gap of the AGNR with rough (ER=25%) edges is larger than that of the AGNR with smooth (ER=0%) edges. As shown in Fig. 5c, for the rough-edged AGNR, the first conductance peak in the positive and negative energy range, respectively, shifts to higher and lower energy. This ER induced conductance gap[43-47] is, however, less significant in the presence of a B-field of B=5 T. Hence, the bandgap reduction due to B-field is larger for AGNR with rough edges, i.e. the shift of the g(E,B) peaks for ER=25% (dotted curve in Fig.

5c) is larger compared to that for the case of ER = 0% (solid curve). This indicates that the atomic disorders at the edges may actually enhance the magnetic sensitivity of the device.

To further analyze these effects, the variation of conductance gap with ER, and the effect of B-field on ER induced conductance gap are summarized in the inset of Fig. 5(b). It can be observed that 1) conductance gap of 15.4 nm wide AGNR increases with increasing ER,[43-47] and 2) the effect of ER on the conductance gap is reduced at higher B-field, i.e. the rate of increase in conductance gap with ER is lower at B=5T compared to that at B=0T. This is because, at higher B-field, the electrons in the GNR are confined by the B-field in addition to the edges, and thus the effect of the edge profiles on conductance gap is less prominent. Therefore, although the current decreases as ER increases under both B=0T and B> 0T, the reduction of the latter such as J(B=5T) is less than that of J(B=0T). As a result, MR of the AGNR is enhanced from 80% to 150% as ER increases from 0% to 35%, as shown in Fig. 5(b). This result indicates that in a more realistic case which incorporates edge imperfections, the MR would actually improve without compromising on the current density.

## IV. Conclusion

In summary, the mechanism of MR effects in AGNRs due to the modulation of bandgap has been proposed. Unlike in the case of conventional graphene-based spin valves, the MR effect does not rely on a change in the relative magnetization orientation of the ferromagnetic leads. Instead, it is an intrinsic effect, which is primarily caused by the subband shifts involved in the formation of n=0 LL under a B-field, resulting in a narrowing of the bandgap. We found that n=3p+1 AGNR group shows the highest sensitivity to a change in the B-field, which translates into the highest MR ratio at smaller B-field values. The MR is, however, suppressed by increasing temperature due to thermal broadening of the Fermi distribution function. We also showed that the operating bias is also one of the key parameters to optimize the MR and driving currents required for device applications. An excessively large bias will cause more electrons to occupy the higher subbands leading to a degradation of MR. Therefore, a careful optimization of these different parameters is necessary to obtain the optimum MR. Finally, we incorporate edge roughness (ER) for a realistic model of the AGNR nanoribbon, and investigate its effect on the carrier transport. Although the presence of ER suppresses the current density, the MR ratio is actually enhanced. This may be explained by the fact that the magnitude of the ER-induced conductance gap diminishes under finite B-fields. This results in a greater reduction of current at zero B-field compared to that at finite B-fields, thus improving the MR ratio.

**ACKNOWLEDGEMENT**

We thank Mr. Kai-Tak Lam for many fruitful discussions. The computations were performed on the cluster of Computational Nanoelectronics and Nano-device Laboratory, National University of Singapore. This work was supported by Agency for Science, Technology and Research, Singapore (A*STAR) under grant number 082-101-0023.

**FIGURE CAPTION**

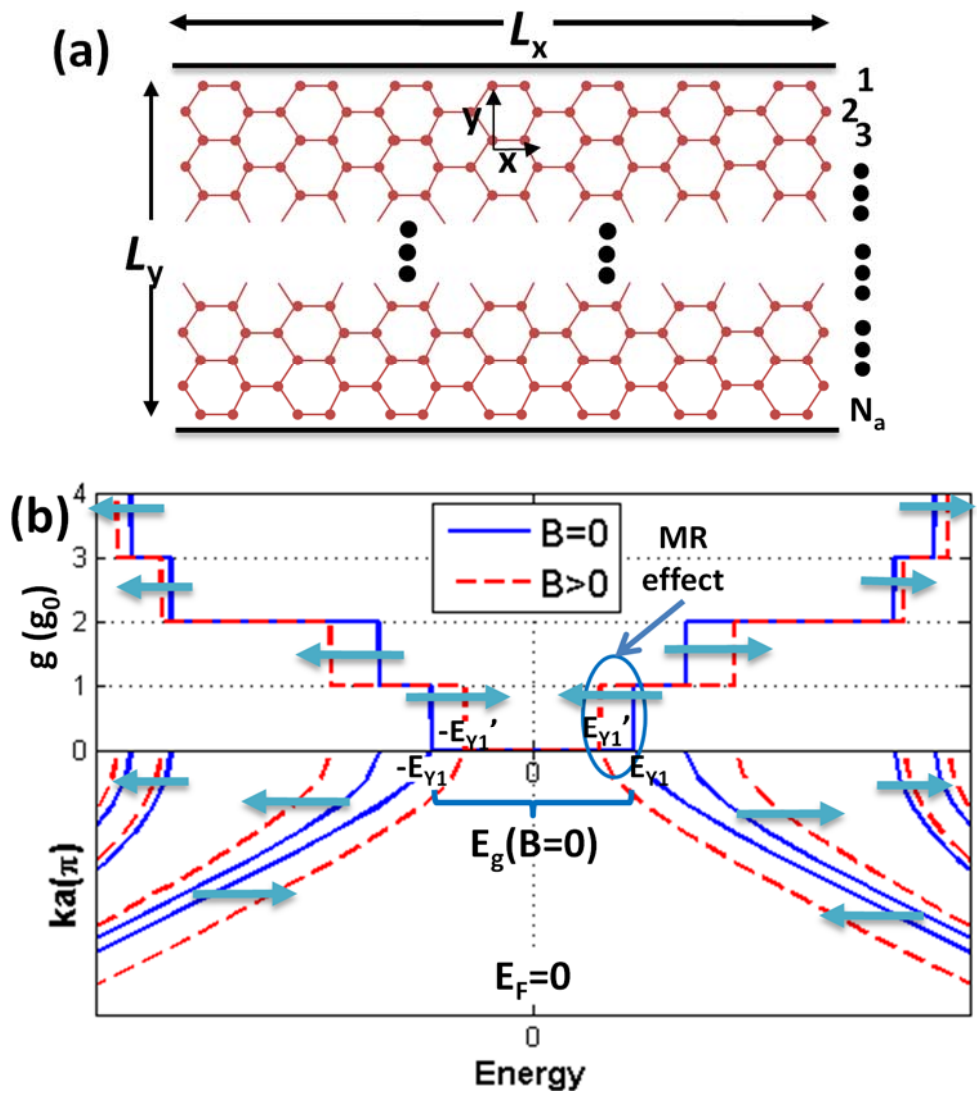

FIG. 1 (color online)  (a) The schematic structure of an AGNR. (b) The E-k band diagram and the zero temperature conductance curve for a typical AGNR. Solid (dotted) curves show the results when a zero (finite) uniform perpendicular magnetic field is applied (B-field). The arrows indicate the shift in curves due to application of B-field. The circle indicates the region within which the MR effect is noticeable.

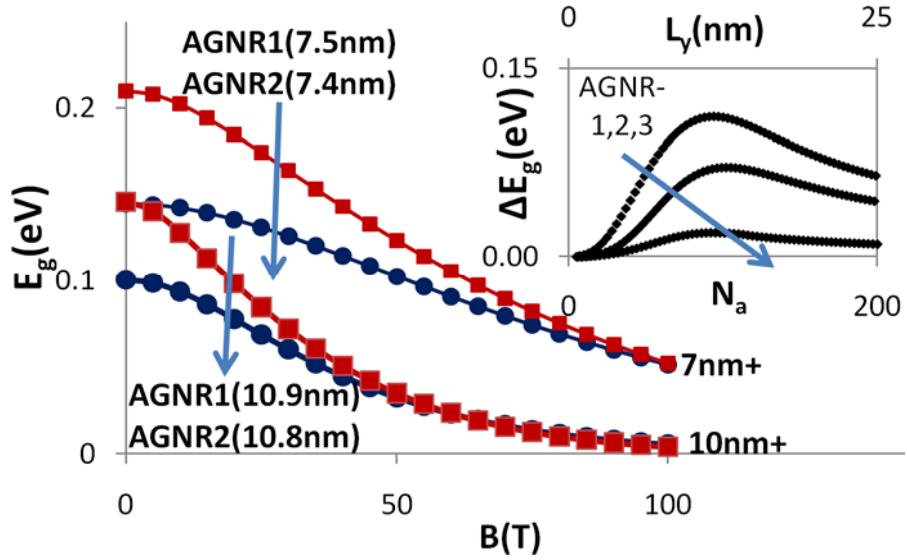

FIG. 2 (color online) The sensitivity of different types of AGNR with applied uniform perpendicular magnetic field (B-field). The band gap ($E_g=2E_{Y1}$) variation of two pairs of AGNR1 and AGNR2 is compared. In both cases, the $E_g$ of AGNR1 drops more drastically compared to AGNR2, implying higher magnetic sensitivity of AGNR1. The inset summarizes the results for magnetic sensitivity of bandgap, $\Delta E_g=E_g(0)-E_g(B)$, where $E_g(0)$ and $E_g(B)$ are bandgaps when B=0T and B>0 (for instance, a value of 50 T is used in this calculation), respectively. Within a small range of $L_y$, AGNR1 is always more sensitive compared to AGNR2/AGNR3. When $L_y$ increases, $\Delta E_g$ increases to a maximum value, and then decreases with further increase in $L_y$.

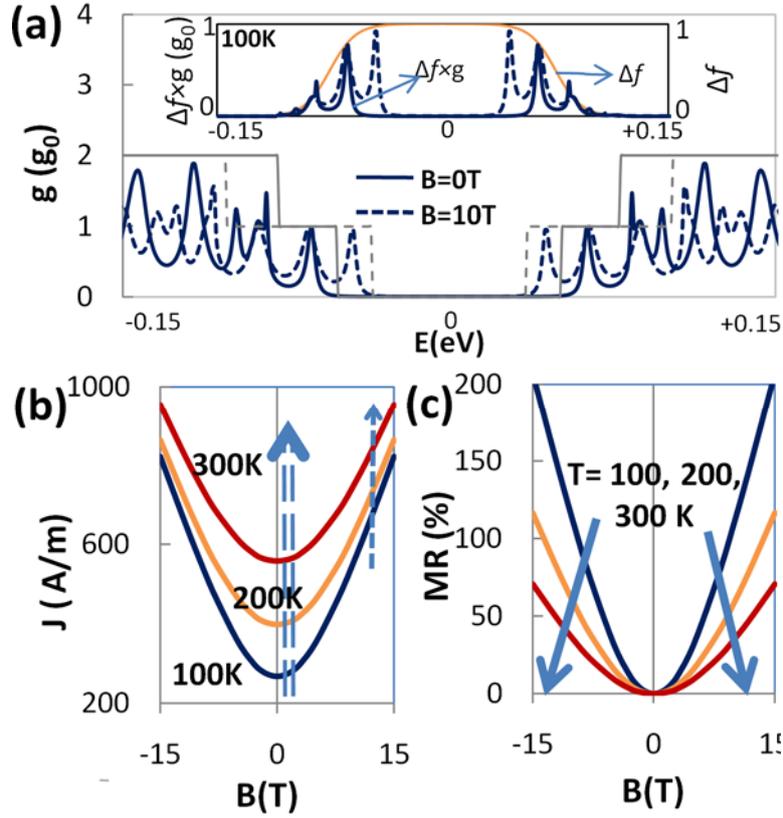

FIG. 3 (color online) (a) The variation of zero temperature conductance, g at different energy levels, E for B=0T and B=10T. The g curve is shifted closer to E=0 at B=10T. The inset shows the variation of $\Delta f \times g$ at T=0. The area below the $\Delta f \times g$ curve is proportional to the total current density, J. This area is larger for the case of B=10T. (b) The variation of J with B-field at T=100K, 200K, and 300K. The increase in J with T is suppressed at larger B. (c) The variation of magnetoresistance (MR) with B-field at T=100K, 200K, and 300K. MR decreases with increasing T. [$L_y$=15.4nm, $V_{SD}$=150mV and $L_x$=65nm.]

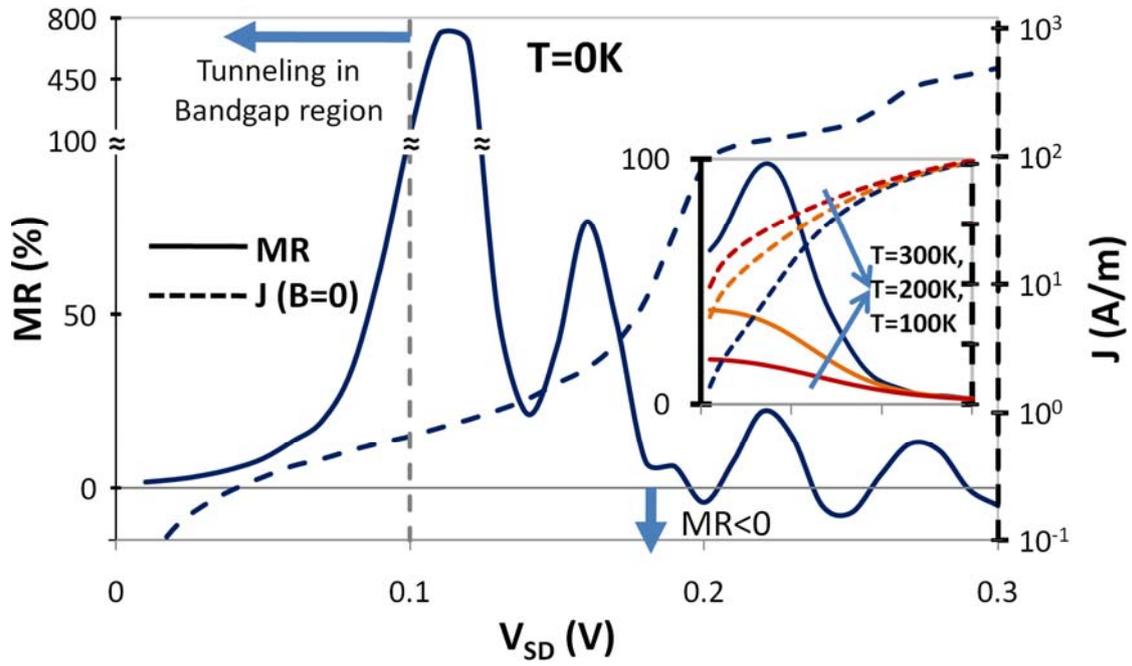

FIG. 4 (color online) The increase in normalized current density, J and variation of MR with increasing $V_{SD}$, indicating the importance of optimizing $V_{SD}$ in order to achieve high MR with sufficient J. [$L_y$=15.4nm, $L_x$=65nm, and T=0K]. In the MR computation, a value of B=5T was used. Inset shows the MR and J for T=100K, 200K, 300K. Unless otherwise stated the axes in the inset are similar to the main figure.

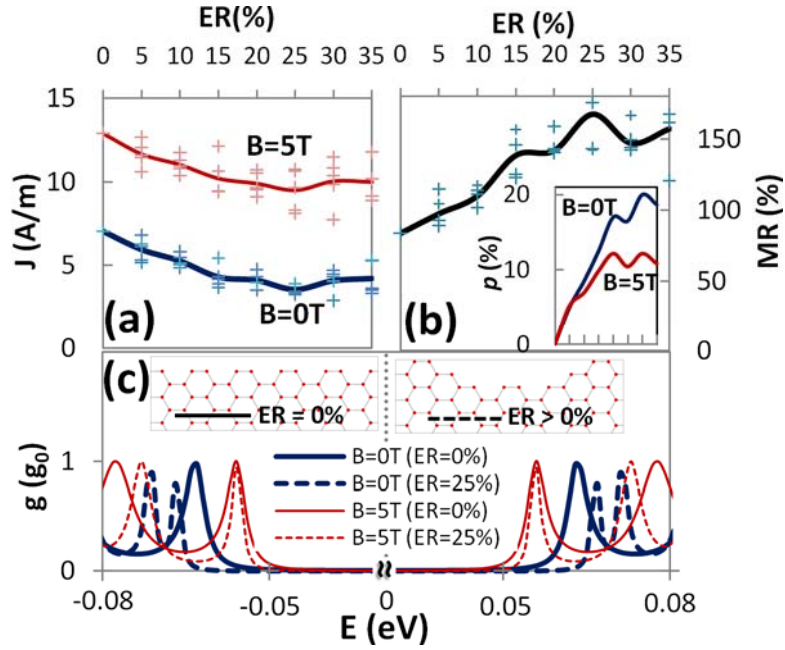

FIG. 5 (color online)  (a) The variation of the current density, J with increasing edge-roughness (ER). (b) The variation of MR with increasing ER. The inset of (b) shows the percentage change in the condutance gap of the device due to ER. The conductance gap increases with ER. However, this rate of increase is suppressed in the presence of a B-field. For each ER, the average of five different samples was taken. (c) The variation of zero temperature conductance, g at different energy levels, E for B=0T (thick lines) and B=5T (thin lines) when ER=0% (solid lines) and ER=25% (dotted lines).  The shift in the curve due to B-field is more significant for the case of ER=25% (dotted lines) compared to the case of ER=0% (solid lines).  The insets of (c) schematically show the structure of a smooth AGNR and that of an AGNR with edge-roughness. [$L_y$=15.4nm, $L_x$=65nm, and T=100K].